\documentclass[journal]{IEEEtran}
\usepackage{latexsym,caption}
\usepackage{amsthm}
\usepackage{amssymb}
\usepackage{setspace}
\usepackage{amsmath}
\usepackage{multirow}
\usepackage{xcolor}
\usepackage[numbers]{natbib}
\usepackage[pdftex]{graphicx}
\usepackage[utf8]{inputenc}

\usepackage[shortlabels]{enumitem}

\begin{document}

\title{Multiscale Data-driven Seismic Full-waveform Inversion with Field Data Study}
%
%
%
\author{Shihang Feng$^{2, *}$, Youzuo Lin$^{1, *}$ and Brendt Wohlberg$^{2}$
\thanks{\textbf{1}: Earth and Environmental Sciences Division, Los Alamos National Laboratory, Los Alamos, NM, 87544 USA.}
\thanks{\textbf{2}: Theoretical Division, Los Alamos National Laboratory, Los Alamos, NM, 87544 USA.}
\thanks{$^{*}$Correspondence to:  S. Feng (shihang@lanl.gov) and Y. Lin (ylin@lanl.gov).}}

\markboth{IEEE Transactions on Geoscience and Remote Sensing}%
{Shell \MakeLowercase{\textit{et al.}}: Bare Demo of IEEEtran.cls for Journals}

\maketitle

\begin{abstract}
Seismic full-waveform inversion (FWI), which uses iterative methods to estimate high-resolution subsurface models from seismograms, is a powerful imaging technique in exploration geophysics. In recent years, the computational cost of FWI has grown exponentially due to the increasing size and resolution of seismic data. Moreover, it is a non-convex problem and can encounter local minima due to the limited accuracy of the initial velocity models or the absence of low frequencies in the measurements. To overcome these computational issues, we develop a multiscale data-driven FWI method based on fully convolutional networks (FCN). In preparing the training data, we first develop a real-time style transform method to create a large set of synthetic subsurface velocity models from natural images. We then develop two convolutional neural networks with encoder-decoder structure to reconstruct the low- and high-frequency components of the subsurface velocity models, separately. To validate the performance of our data-driven inversion method and the effectiveness of the synthesized training set, we compare it with conventional physics-based waveform inversion approaches using both synthetic and field data. These numerical results demonstrate that, once our model is fully trained, it can significantly reduce the computation time, and yield more accurate subsurface velocity models in comparison with conventional FWI.
\end{abstract}

\IEEEpeerreviewmaketitle

\section{Introduction}

Accurately and efficiently characterizing subsurface geology is crucial for various applications, such as energy exploration, civil infrastructure, groundwater contamination and remediation, etc. The standard approach to obtaining such a characterization is via computational seismic imaging, which involves reconstructing an image of subsurface structures from measurements of natural or artificially produced seismic waves~\cite{virieux2009overview}. Inspired by recent successes in applying deep learning to computer vision and medical problems, deep-learning-based data-driven methods had been applied to seismic imaging problems. Several encoder-decoder networks have been developed to reconstruct the subsurface structure from seismic data~\cite{Wu-2019-InversionNet, li2019deep, yang2019deep, Zhang-2020-Data}. Those deep-learning models are end-to-end, meaning that they use the seismic waveform data as the input and directly output its corresponding subsurface structure. Once those models are fully trained, the inversion procedure is computationally efficient. However, a significant weakness of these data-driven methods is their weak generalization ability, which hinders the wide  application of data-driven seismic imaging approaches to field data~\cite{Zhang-2020-Data}. 

Weak generalization is a common challenge facing all deep-learning applications. It means the predictive models trained in a specific dataset cannot perform well when applying to an out-of-distribution dataset. To improve the generalization ability, novel models have been developed to incorporate physics laws and prior information (such as geometric rules, symmetries or other relevant constraints) into the deep learning models~\cite{Gomez-2020-Physics, Sun-2020-theory}. Alternatively, the robustness and generalization ability of deep learning models can be improved by acquiring more labeled data. However, neither of these solutions is straightforward for seismic imaging. The current state-of-the-art physics-based~(theory-guided) FWI approaches can provide limited constraints with respect to the governing physics. Furthermore, it can be extremely difficult and expensive to collect real subsurface structure models and their corresponding seismic measurements, which results in training sets with limited representativeness. To overcome the weak generalization issue, we explore the possibility of enriching the training set and incorporating critical physics phenomena in our predictive model. 

 
 Our idea is inspired by the artistic style transfer problems from the computer vision community, the goal of which is to transfer the art style of one painting to another image by minimizing the style loss and the content loss based on features extracted from a pre-trained convolutional neural network (CNN)~\cite{johnson2016perceptual, gatys2015neural}. Those tools therefore provide us with approaches to bridge images from two different physical domains. Specifically, subsurface structure models represent the geophysical properties in 2D, which can be also viewed as images of a certain physical property.~\cite{ovcharenko2019style} firstly implemented an iterative style transfer approach in generating the practical velocity models. The style transfer is regarded as an iterative optimization task and a set of subsurface structure models used as the content image. Their  method strongly relies on domain knowledge to generate the content images, which in turn limits the variability of the training set. In addition, the iterative optimization is not computational effective, which significantly limits the size of the training set. In this paper, we employ the one developed by~\cite{johnson2016perceptual} due to its efficiency in solving the optimization. A feed-forward style transfer network can be run in real-time after training, so that it is feasible to generate numerous synthetic art images efficiently. It converts a large volume of existing natural images into subsurface structure models with pre-determined geologic styles. In such a manner, our method can generate a large number of synthetic subsurface velocity models with sufficient variability. That in turn not only helps our data-driven models to learn the governing physics (forward model) of the problem through training, but also yields high generalization ability due to the richness of the data representativeness.

Incorporation of critical physics into neural network structures also plays an important role in improving the robustness of predictive models~\cite{Gomez-2020-Physics, Raissi-2019-Physics, raissi-2017-physicsI,alkhalifah2020wavefield}. Unlike conventional FWI, where the relationship between seismic data and velocity model is governed by the wave-equation, data-driven FWI methods learn a correspondence from seismic data directly to subsurface structure. The governing physics of the forward modeling is represented implicitly in the training dataset~\cite{wang2020velocity}. On the other hand, the propagation of seismic wave is a complex physical phenomenon, which consists of  different waves, such as reflection waves, transmission waves and direct waves, etc. Each of them follows different wavepaths and propagation mechanisms. To account for the complex physics and better explain the various wave phenomena, we develop a multiscale inversion strategy. It is worthwhile to mention that multiscale techniques have been widely used in convectional physics-based FWI approaches~\cite{bunks1995multiscale, alkhalifah2014tomography, alkhalifah2016full, feng2019transmission+, liu2019multiscale} to handle different waves and preventing local minima. With this strategy incorporated, our data-driven FWI is able to separately invert transmission and reflection waves in the seismic data. Particularly, the predicted results from the low-frequency components can be used as the initial guess for the high-resolution inversion, which significantly improves the overall inversion accuracy.

With the aforementioned two network modules~(i.e., style-transformed network and multiscale data-driven inversion network), we first train a feed-forward style transfer network to generate numerous synthetic velocity models. Those velocity models and their corresponding seismic data are then utilized to train our multiscale data-driven FWI networks, called ``Multiscale InversionNet''. Once the network is fully trained, the model can effectively and efficiently invert the seismic data to obtain velocity models. 


This paper is organized in seven sections. After the introduction, the second section briefly reviews the related work. The third section
presents the theory of image style transfer and seismic full-waveform inversion. Our proposed methodologies of building practical velocity models and multiscale InversionNet are introduced in the fourth section. The fifth section shows the numerical results with both synthetic and field data. A discussion of our Multiscale InversionNet is presented in the sixth section. Finally, the conclusion is given in the last section. 

\section{Related Works}

\subsection{Physics-based Full-waveform Inversion}

Building the velocity models is an important task in the study of subsurface characterization. Full-waveform inversion techniques~\cite{brossier2015velocity} provide superior solutions by modeling the wave propagation in the subsurface. The problem of FWI is ill-posed, usually without a unique solution~\cite{virieux2009overview}, making the initial guess of the solution space and low-frequency information of the data space essential~\cite{alkhalifah2014full}. Multi-scale FWI methods are developed to mitigate the non-linearity issue of FWI~\cite{bunks1995multiscale,sirgue2004efficient,boonyasiriwat2009efficient}. Once the smooth or low-wavenumber velocity structures are obtained, the high-wavenumber structures are reconstructed using high-frequency data. However, the convergence to geological meaningful models is not always guaranteed if the initial guess is far away from the true model~\cite{altheyab2015reflection,datta2016estimating}. Waveform inversion of the early arrival data reconstructs the low-wavenumber content in the shallow subsurface~\cite{shen2010near,yu2014application} while the updating of the low-wavenumber in the deep part of the subsurface can be challenging. Reflection waveform inversion~\cite{chi2015correlation,guo2017elastic,chen2020multiscale} uses a prior reflectively model and builds the reflection wavepaths to update the low-wavenubmer content in the deep region. By combining the early arrival and the reflection waves, a joint FWI method~\cite{zhou2015full} can update the smooth content of the velocity model and the result can be used as the initial guess for the classical FWI.

Full-waveform inversion builds on the forward modeling solutions. Current numerical optimization techniques for solving FWI require hundreds or even thousands of times of forward modeling. Existing forward modeling techniques are based on by finite difference~\cite{alford1974accuracy}, finite-element~\cite{choi2008two}, spectral wavefield extrapolation~\cite{wu2014optimized}, pseudo-spectral~\cite{kosloff1982forward} and spectral-element~\cite{komatitsch2005spectral} methods. All those techniques are non-linear and computationally intensive. Furthermore, with the explosive growth in data volumes due to developments in seismic acquisition technology~\cite{poole2018thef}, solving FWI problems for large-scale data set becomes computationally prohibitive or even infeasible. To mitigate those technical limitations, data-driven FWI techniques have recently been developed.


\subsection{Data-driven Full-waveform Inversion}
\label{sec:datadriven}

Various machine-learning-based FWI techniques have been recently developed. Broadly speaking, most of the existing data-driven techniques fall into the following three major categories: the end-to-end techniques, the low-wavenumber learning techniques,  and the theory-guided data-driven techniques. Due to a large volume of work out of each category, we can only select those that are most representative to its own kind. A more thorough review on this topic can be found in~\cite{Adler2021Deep}. 

The end-to-end techniques attract the most interest due to its direct tie to the universal approximating nature of machine learning~\cite{Deep-2018-Araya,Wu-2019-InversionNet,kazei2019mapping,li2019deep,yang2019deep,wang2020velocity, Data-2020-Zhang}. The commonality among all those end-to-end approaches is that a FWI inversion operator is learned using neural network to obtain the velocity models directly from seismic data. \cite{Deep-2018-Araya} is the first of its kind, which is built on fully connected neural network. InversionNet~\cite{Wu-2019-InversionNet} introduced an encoder-decoder structure with conditional random field. Instead of using common-shot gather, \cite{kazei2019mapping} exploited and developed feed-forward network to take advantage of common midpoint gathers. SeisInvNet~\cite{li2019deep} introduced a mixed loss function in the training of the encoder-decoder network to optimize the quality of the reconstructed velocity models. The skipping layers are applied in the encoder-decoder architectures to construct velocity models with salt bodies~\cite{yang2019deep}. ModifiedFCN~\cite{wang2020velocity} extended the data-driven FWI to cross-well data in the hydrocarbon production site. VelocityGAN~\cite{Data-2020-Zhang} employed generative adversarial network~(GAN) to solve FWI and further improved the model robustness using transfer learning strategy. 

The low-wavenumber information in the data is critical in obtaining accurate velocity models when solving FWI. On the other hand, extraction of those information is also technically challenging. The low-wavenumber learning techniques are developed to address this challenge~\cite{Deep-2019-Ovcharenko, Extrapolated-2020-Sun, Progressive-2021-Hu}. \cite{Deep-2019-Ovcharenko} designed a deep CNN to synthesize low-frequency representation of the a shot gather provided with the high-frequency components. \cite{Extrapolated-2020-Sun} developed a CNN model to extrapolate the low-frequency information trace-by-trace in the time domain. \cite{Progressive-2021-Hu} leveraged active learning to extrapolate missing low-wavenumber component using both beat tone and seismic waveform data. 

One potential issue associated with the aforementioned approaches is the weak generalization ability. To enhance the generalization, an interesting direction is to develop a hybrid model by combining machine learning and governing physics. That leads to the theory-guided data-driven techniques~\cite{Gomez-2020-Physics, theory-2020-Sun,Physics-2020-Ren}. An adaptive data augmentation technique was developed in \cite{Gomez-2020-Physics} to improve the data representativeness and incorporate physics knowledge. \cite{theory-2020-Sun} represented the wave equation using recurrent neural network and that would allow physics knowledge being incorporated during training procedure. \cite{Physics-2020-Ren} employed a different strategy to discretize the wave equation for the purpose of building physics-guided neural network structure.

 \subsection{Velocity Model Constructions for Data-driven Inversion}

A high quality training dataset with sufficient representativeness is the foundation for obtaining a robust predictive model~\cite{Schat-2020-data}. Particularly for seismic imaging, a training set with high representativeness should not only account for geological variability but also be compatible with domain knowledge and intrinsic physics rules. 

To account for the requirement of geologic knowledge in the training set, the most widely-used method is to fix the types of geological structures and randomize the properties of these structures.~\cite{li2019deep} and~\cite{fabien2020seismic} considered 2D laterally homogeneous isotropic layers in the velocity models. ~\cite{Deep-2018-Araya}, \cite{yang2019deep} and \cite{sun2021physics} built a training set by inserting a single salt body with an arbitrary shape and location in background models with a series of laterally layers. ~\cite{Wu-2019-InversionNet} used the models with flat and curved subsurface layers as the benchmarks.~\cite{Wu-2020-Building} developed a workflow to automatically build diverse structure models with realistic folding and faulting features.~\cite{ren2021building} proposed a framework to generate 3D velocity models with folding layers, faults and salt bodies.~\cite{kazei2019mapping} presented a workflow to generate pseudo-random subsurface models by flipping, distorting and cropping the known guiding model. These method relies on the assumptions on the structures in the velocity models, e.g., layering, faults, salt bodies and so on, therefore, producing unsatisfactory results when applying to different sites.
 
To increase the generalization of the neural network,~\cite{wang2020velocity} introduced the natural images in the velocity model construction to enlarge the diversity of the training set. However, the lack of the geology information limited its application in the field data.~\cite{ovcharenko2019style} applied a style transfer method to obtain velocity models from limited content images to output velocity models with pre-determined geologic styles, but the iterative optimization is computationally expensive regarding a large training set.

\section{Theory}
\subsection{Seismic Full-waveform Inversion}
The forward model of our problem is the acoustic-wave equation, which is given by 
\begin{equation}
\nabla^2p(\mathbf{r},t)-\frac{1}{c^2(\mathbf{r})}\frac{\partial ^2p(\mathbf{r}, t)}{\partial t ^2}=s(\mathbf{r},\, 
t)
\label{eq:Forward}
\end{equation}
where $c (\mathbf{r})$ is the velocity at spatial location 
$\mathbf{r}$, $\nabla^2=\left(\frac{\partial^2}{\partial{x^2}}+\frac{\partial^2}{\partial{z^2}}\right)$ is the Laplacian operator in 2D Cartesian coordinates, $s(\mathbf{r},\, 
t)$ is the source term, $p(\mathbf{r}, t)$ is the pressure data, and $t$ represents time. To simplify the expression, we rewrite the forward modeling problems in Eq.~\eqref{eq:Forward} as
\begin{equation}
\mathbf{d}=f(\mathbf{m}) \;,
\end{equation}
where  $\mathbf{d}$ is the pressure data for the acoustic case, $f$ is the forward acoustic-wave modeling operator, and $\mathbf{m}$ is the model parameter vector, which is the compressional-~(P-) velocities.  
\subsubsection{Physics-based Full-waveform Inversion}
In physics-based full-waveform inversion, the objective function is often given by
\begin{equation}
l_{\mathrm{FWI}}=\frac{1}{2}\sum_{s,g}{||\mathbf{d}_{s,g}^{pre}-\mathbf{d}_{s,g}^{true}||^2},
\label{eq:fwiloss}
\end{equation}
where $\mathbf{d}^{pre}_{s,g}$ and $\mathbf{d}^{true}_{s,g}$ are the predicted and observed data at the locations of receivers $\mathbf{g}$ for sources $\mathbf{s}$. The model is gradually updated using gradient-based optimization methods to approximate $f^{-1}$, which are computationally expensive.

\subsubsection{Data-driven Full-waveform Inversion}
Unlike the physics-based full-waveform inversion, which preforms the inversion using an iterative method, data-driven seismic inversion obtains an approximation of $f^{-1}$ by training an encoder-decoder network~\cite{Wu-2019-InversionNet}. It achieves the data-driven FWI by regressing seismic data $\mathbf{d}$ to velocity model $\mathbf{m}$ directly. In this work, we employ InversionNet as the building block of our network structure. More details of InversionNet can be found in our previous  paper~\cite{Wu-2019-InversionNet}. For readers who are interested in broad data-driven FWI approaches, please refer to those work provided in Section~\ref{sec:datadriven}.

\begin{figure*}[ht]
\centering
\includegraphics[width=1.70\columnwidth]{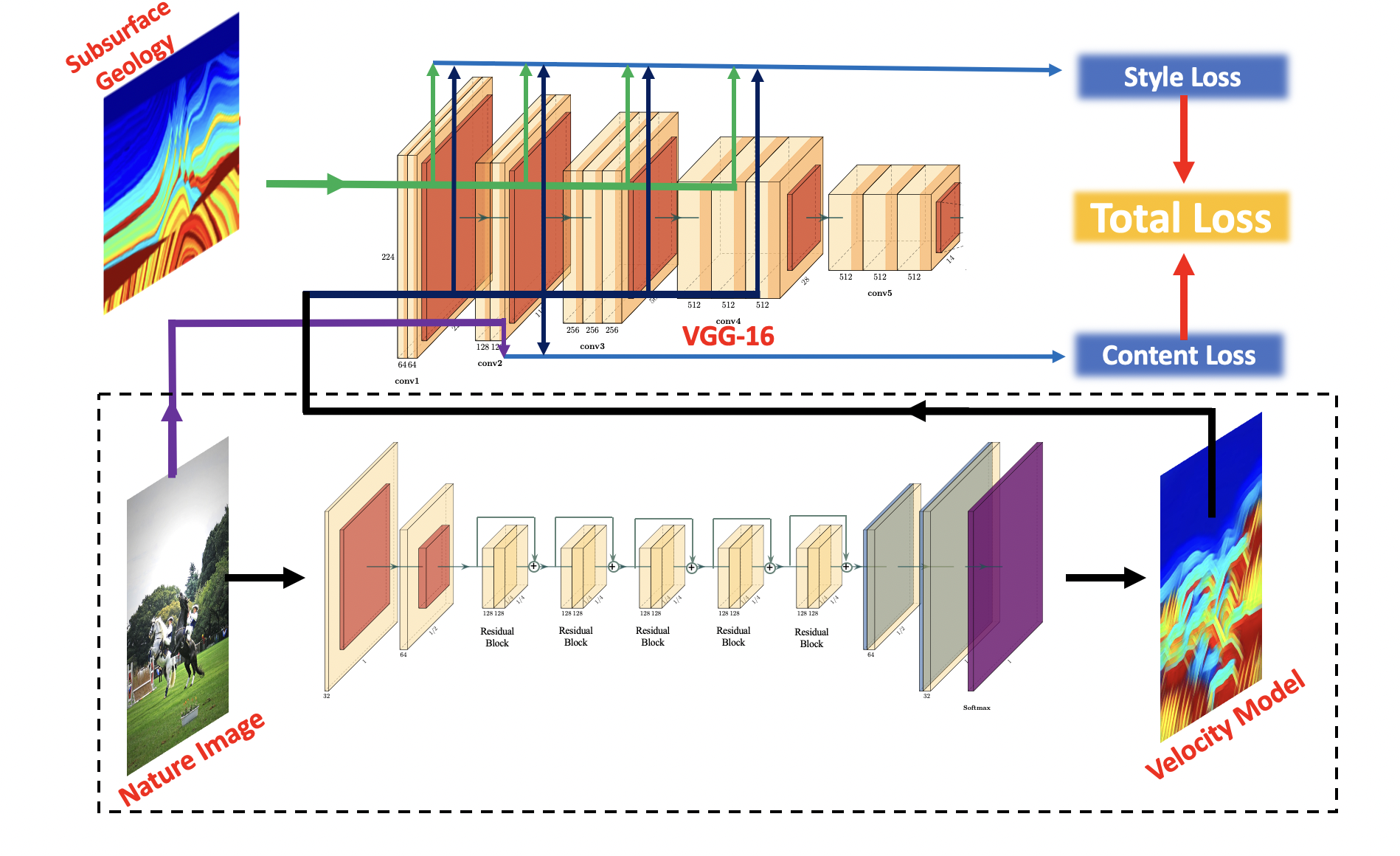}
\caption{A schematic illustration of our style-transform network. The image transform network is an encoder-decoder network, which is trained to transform natural image into a velocity perturbation with style loss and content loss. A pre-trained VGG-16 network is used to extract the feature models from the natural images and the subsurface geology, it is fixed during the training procedure. Style loss is defined as the Gram matrix difference between the feature models of the natural images and the subsurface geology in $\mathrm{relu1}\_\mathrm{2}$, $\mathrm{relu2}\_\mathrm{2}$, $\mathrm{relu3}\_\mathrm{3}$ and $\mathrm{relu4}\_\mathrm{4}$ layers. Content loss is the Mean squared error (MSE) loss between the feature models of the natural images and the subsurface geology in $\mathrm{relu2}\_\mathrm{2}$ layer.}
\label{fig:style_net}
\end{figure*}

\subsection{Image Style Transfer}
Image style transfer can be defined as finding a composite image $y$ whose style is similar to a style image $y_s$ and content is similar to a content image $y_c$.  Two images are similar in style if their low-level features extracted by a trained classifier have similar Gram matrix $G_j(x)_{mn}$~\cite{gatys2015neural,ovcharenko2019style}
\begin{equation}
G_j(x)_{mn}=\sum_{p}\phi_j{(x)}_{mp}\phi_j{(x)}_{np} \;,
\end{equation}
where  $\phi_j(x)$ is the activations at the $j$th layer of the network $\phi$ for the input image $x$ and the Gram matrix $G_j(x)_{mn}$ is the inner product between the vectorized feature maps $m$ and $n$ in layer $j$. $p$ is the index of the vectorized feature maps in layer $j$. The style loss is defined as 
\begin{equation}
l_\mathrm{style}=\sum_{j\in{S}}\frac{1}{U_j}||G_j(y)-G_j{(y_s)}||^2 \;,
\end{equation}
where $S$ is a set of layers used in style reconstruction, $U_{j}$ is the total number of units in layer $j$ and $y_s$ is the style image and $y$ is the composite image. Two images are similar in content if their high-level features extracted by a trained classifier are close. The content loss is defined as below
\begin{equation}
l_\mathrm{content}=\sum_{j\in{C}}\frac{1}{U_j}||\phi_j{(y)}-\phi_j{(y_c)}||^2 \;,
\end{equation}
where $C$ is a set of layers used in content reconstruction and $y_c$ is the content image. The real-time style transfer system is shown in Figure~\ref{fig:style_net}, where the right side is the calculation of the loss function of the network. The overall loss is defined as
\begin{equation}\label{eq:loss}
l_{trans}{=\alpha_\mathrm{style}\,l_\mathrm{style}+\alpha_\mathrm{content}\,l_\mathrm{content}} \;,
\end{equation}
where $\alpha_\mathrm{style}$ and $\alpha_\mathrm{content}$ are the weights for style and content reconstruction. By defining the loss function in Eq.~\eqref{eq:loss}, the image transform feed-forward networks are trained to solve the optimization problem.

\section{Methodology}

\subsection{Building Physically Practical Velocity models}

The current data-driven seismic FWI approaches rely heavily on the pre-generated simulations. However, the unavoidable discrepancy between simulations and field data causes the limitation of the representiveness of the training dataset, which hinders its application to field data. In order to bridge the gap between simulation and field data, we expect a high-quality training dataset should consist of a large volume of subsurface velocity models with sufficient variability in order to represent the complex geology in various scenarios. To our best knowledge, there is no such a dataset existing for training seismic FWI problems. To overcome this data challenge, we develop a data generation approach that is capable of synthesizing a large volume of physically practical subsurface velocity models efficiently. Specifically, our approach is built on natural image dataset~(COCO dataset~\cite{lin2014microsoft} in this work), taking advantage of its large sample size, high image quality, and varying visual perception. We develop a domain adaptation technique to transfer natural images from COCO dataset to subsurface velocity models. Our data-generation technique can produce a large number of synthetic subsurface velocity models that is consistent with the subject matter expertise. 

Inspired by the work of~\cite{johnson2016perceptual}, we design a neural network to generate subsurface velocity models as shown in Figure~\ref{fig:style_net}. The inputs of our network include the content natural image and the style image. We convert the content natural image into a subsurface structure model, which contains the geologic features learned from the style image. Particularly, in this network architecture, we use the $\mathrm{relu1}\_\mathrm{2}$, $\mathrm{relu2}\_\mathrm{2}$, $\mathrm{relu3}\_\mathrm{3}$ and $\mathrm{relu4}\_\mathrm{3}$ layers in VGG16 network~\cite{Simonyan-2015-Deep} for style reconstruction and the $\mathrm{relu2}\_\mathrm{2}$ for content reconstruction.

\begin{figure}[ht]
\centering
\includegraphics[width=1\columnwidth]{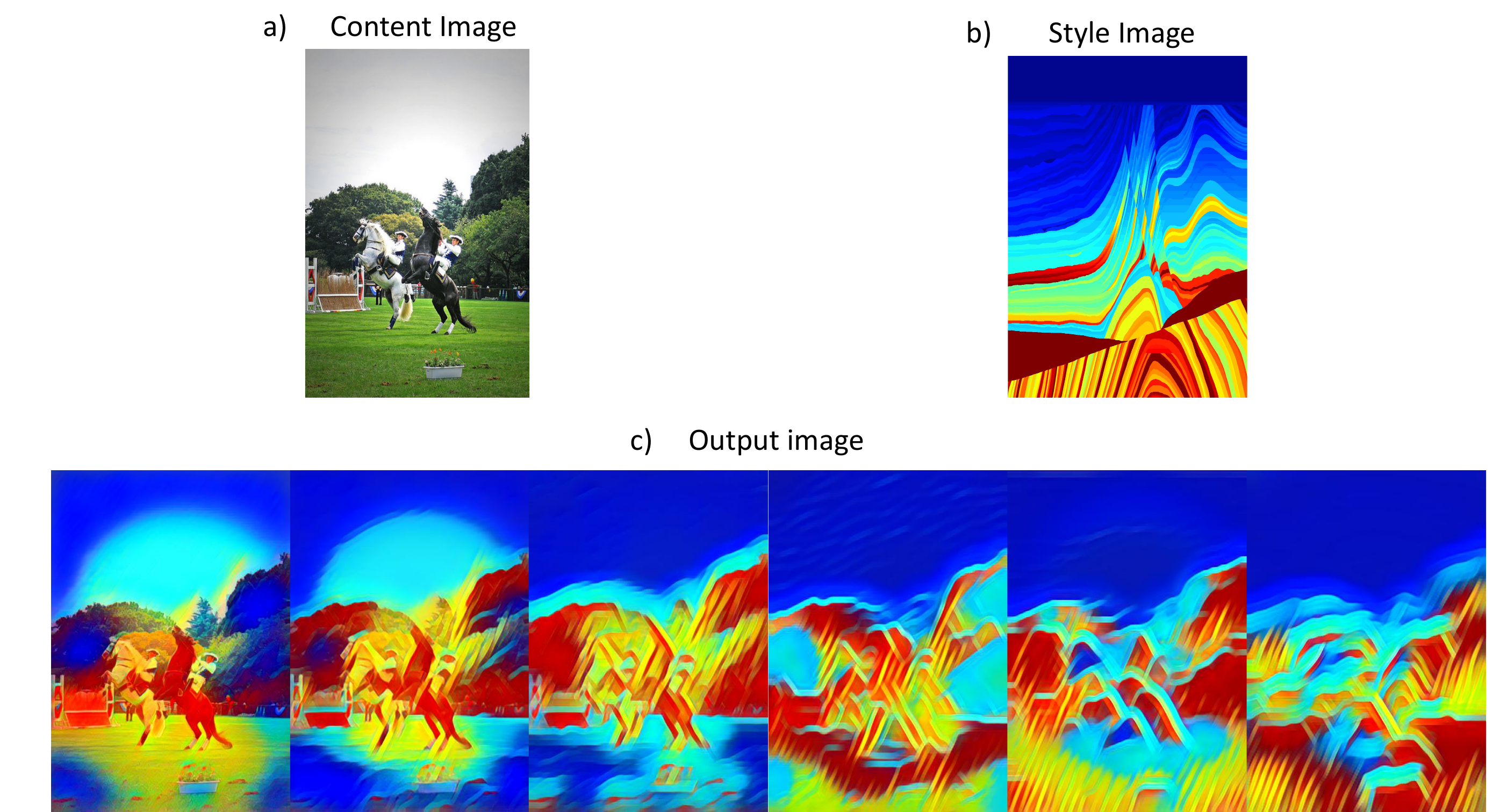}
\caption{a) Content image, b) style image and c) output style transfer images with style weight = 1e9, 3e9, 5e9, 8e9, 1e10 and 5e10.}
\label{fig:style_weight}
\end{figure}

The results with different style weights are shown in Figure~\ref{fig:style_weight}. As the style weight increases, the composite image contains more geological features and becomes more similar to the subsurface structure. 

The color distributions in natural images are different from the velocity distributions in the style image. Since the composite images obtain their content from the natural images, there are clear differences between the composite images and the style image.

\begin{figure}[ht]
\centering
\includegraphics[width=1\columnwidth]{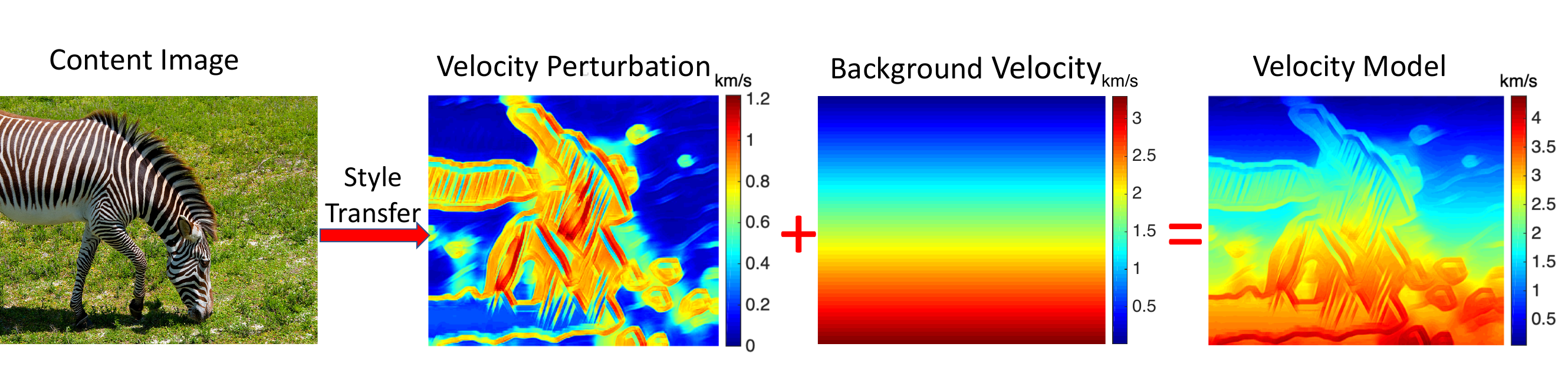}
\caption{The workflow of creating practical velocity models. The content images are transferred to velocity perturbation model using the trained style transfer network. The 1D linear-increased background velocity is then imposed to compose practical velocity models.}
\label{fig:velocity}
\end{figure}

To mimic the real subsurface velocity models, the composite image obtained from style transfer is converted to a single-channel gray-scale image and normalized to a velocity perturbation model. Next, a 1D velocity model with linearly increasing value is utilized as the background velocity. The composite velocity model as shown in Figure~\ref{fig:velocity} is obtained by combining these two models using following equation:
\begin{equation}\label{eq:velocity}
v_{com}=\beta_{pert}v_{pert}+(1-\beta_{pert})v_{back},
\end{equation}
where $v_{com}$, $v_{pert}$ and $v_{back}$ are the composite velocity model, the velocity perturbation model and the background velocity model, respectively. $0.1 < \beta_{pert} < 0.3$ is the weight of the velocity perturbation model. We therefore synthesize a physically meaning subsurface velocity model, which inherits  geological features from the style image. 

\subsection{Multiscale InversionNet}

The kinematics of seismic wave propagation is complex, which makes full-waveform inversion a complicated problem. The multiscale methodology is able to break the complex inversion into a series of simpler inversion procedures. Such methodology has been applied in physics-based full-waveform inversion to mitigate the local minimum problem by proceeding the seismic data from low to high frequency. In this section, we will study how to apply such a multiscale idea in the data-driven full-waveform inversion. 

Built on our previous work of InversionNet~\cite{Wu-2019-InversionNet}, our Multiscale InversionNet consists two modules: a Low-resolution InversionNet and a High-resolution InversionNet. Particularly, the Low-resolution InversionNet is used to invert the low-frequency components of the velocity models, and the High-resolution InversionNet is applied to reconstruct the high-frequency components of the velocity models.  

\subsubsection{Low-resolution InversionNet}
\begin{figure*}[ht]
\centering
\includegraphics[width=1.2\columnwidth]{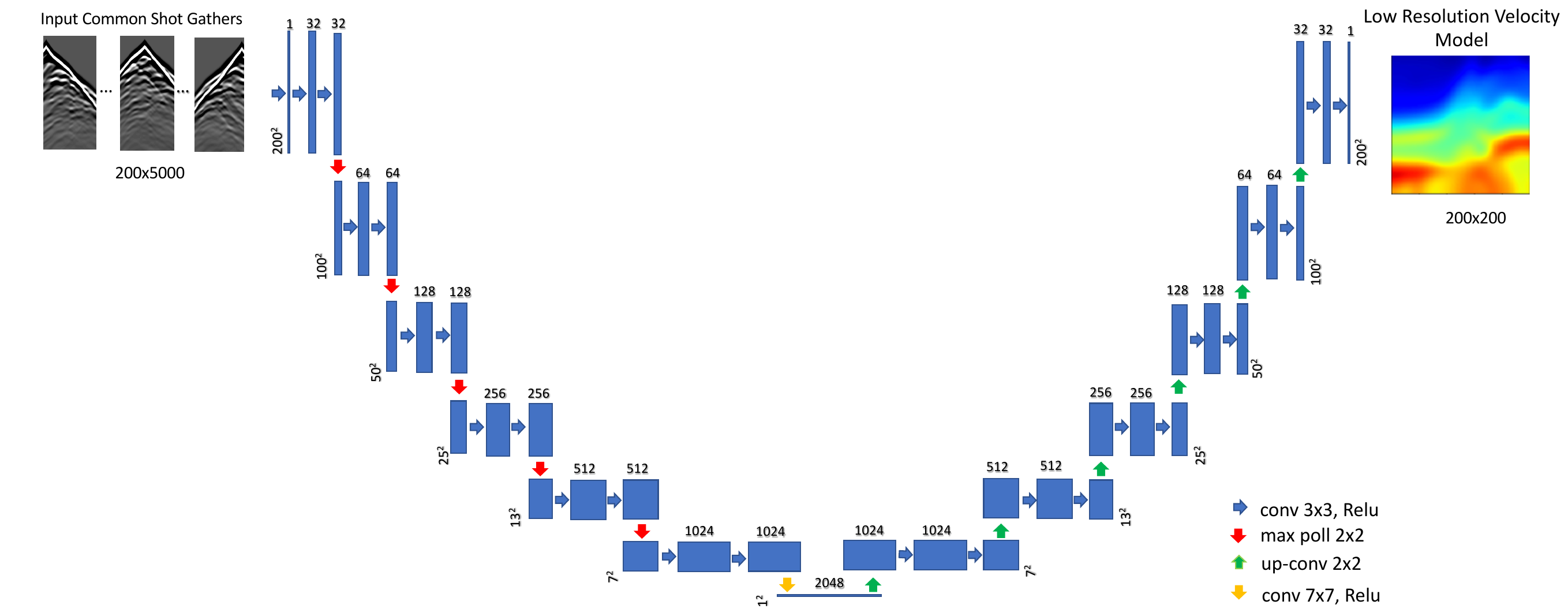}
\caption{A schematic illustration of Low-resolution InversionNet, which is an encoder-decoder network. The encoder is primary built with convolutional layers, which extract the high-dimensional feature models from the seismic data. The decoder projects the extracted feature models into velocity models through a set of deconvolutional layers. }
\label{fig:fcn_full_net}
\end{figure*}

To invert the low-frequency component of the velocity model, we design the architecture of the Low-resolution InversionNet as shown in Figure~\ref{fig:fcn_full_net}. We  choose the $\ell_2$ loss function as our optimality criterion:
\begin{equation}\label{eq:loss_fcn}
{l_{inv\_low}}=\frac{1}{N}\sum^{N}_{i=1}||\mathbf{m}_i^\mathrm{pre\_{low}}-\mathbf{m}_i^\mathrm{true\_{low}}||^2 \;,
\end{equation}
where $\mathbf{m}_i^\mathrm{pre\_{low}}$ and $\mathbf{m}_i^\mathrm{true\_{low}}$ are the low-resolution velocity model predicted by network and the ground truth with $i$th training sample, respectively. The variable $N$ is the total number of the training samples. For a more detailed discussion of loss function selection, please refer to our earlier work \cite{Wu-2019-InversionNet}. 

\subsubsection{High-resolution InversionNet}

\begin{figure*}[ht]
\centering
\includegraphics[width=1.2\columnwidth]{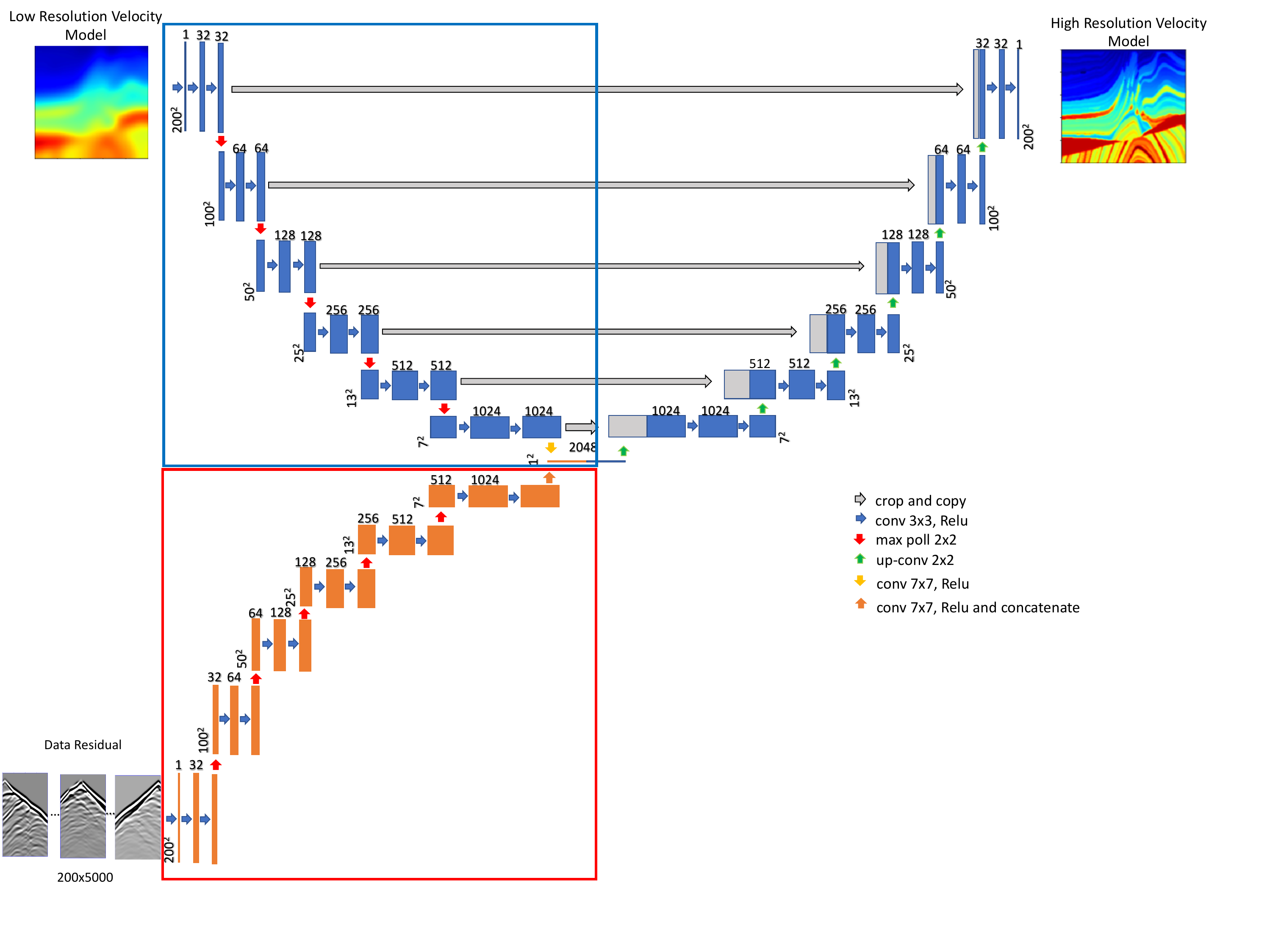}
\caption{A schematic illustration of High-resolution InversionNet.  The encoder contains two parts, a model part and a data part. The model part extracts the feature models from the low-resolution velocity model through convolutional layers and it is directed connected with the decoder through skip connections. The data part extracts the feature models from the data residual. The extracted feature models from the model part and the data part are combined and fed into the decoder to reconstruct the high-resolution velocity model.}
\label{fig:fcn_full_net2}
\end{figure*}

The purpose of our High-resolution InversionNet is to refine the low-frequency velocity component learned via the Low-resolution InversionNet by accounting for reflection wave from the data. To achieve this, we design a different encoder-decoder network architecture as shown in Figure~\ref{fig:fcn_full_net2}. The encoder consists of two parts: a model part and a data part. The model part, as shown in blue box~(Figure~\ref{fig:fcn_full_net2}), is a U-net~\cite{UNet-2015-Ronneberger} like encoder that incorporates the low-frequency information of the velocity model into the network. The input is the predicted low-resolution velocity model from Low-resolution InversionNet. The second part, as shown in red box~(Figure~\ref{fig:fcn_full_net2}), is an encoder that adds data information in the network. The input is the data residual calculated as
\begin{equation}\label{eq:data_diff}
{\mathbf{d}_i^\mathrm{diff}}=\mathbf{d}_i^\mathrm{pre}-\mathbf{d}_i^\mathrm{true} \;,
\end{equation}
where $\mathbf{d}_i^\mathrm{pre}$ and $\mathbf{d}_i^\mathrm{true}$ are the predicted data generated from low-resolution velocity model and the observed data with $i$th training sample, respectively. 
The loss function can be define as a $\ell_2$ loss function,
\begin{equation}\label{eq:high_loss_fcn_l2}
{l_{inv\_high}}=\frac{1}{N}\sum^{N}_{i=1}\|\mathbf{m}_i^\mathrm{pre\_{high}}-\mathbf{m}_i^\mathrm{true\_{high}}\|_2^2 \;,
\end{equation}
where $\mathbf{m}_i^\mathrm{pre\_{high}}$ and $\mathbf{m}_i^\mathrm{true\_{high}}$ are the low-resolution velocity model predicted by network and the ground truth with $i$th training sample, respectively. A loss function using $\ell_1$ norm can be derived, similarly. We include it in the Supplemental Material. 

\subsubsection{Training Strategy}
Training both the model- and data-parts in the High-resolution InversionNet as shown in Figure~\ref{fig:fcn_full_net2} can be technically challenging due to the different physical properties of inputs. A straight-forward training strategy to train our High-resolution InversionNet simultaneously can lead to an unbalanced convergence, meaning that one part of the network would dominate the training procedure over the other part. To mitigate this issue, we design a two-step alternate training scheme:

\begin{enumerate}[i.]
\item Fix the parameters in the red box and update the parameters in the blue box and the decoder so that the low-frequency information of the velocity model can be leveraged in the learning process.

\item Fix the parameters in the blue box and update the parameters in the red box and the decoder. Similar to conventional FWI, the data residual is used to update the velocity model to reconstruct the high frequency information.
\end{enumerate}
We alternate the training procedure between Steps~1 and 2. Once the training is completed, our High-resolution InversionNet can be used to infer subsurface velocity model with an initial guess from the Low-resolution InversionNet and the data differences.
\begin{figure*}[ht]
\centering
\includegraphics[width=1.7\columnwidth]{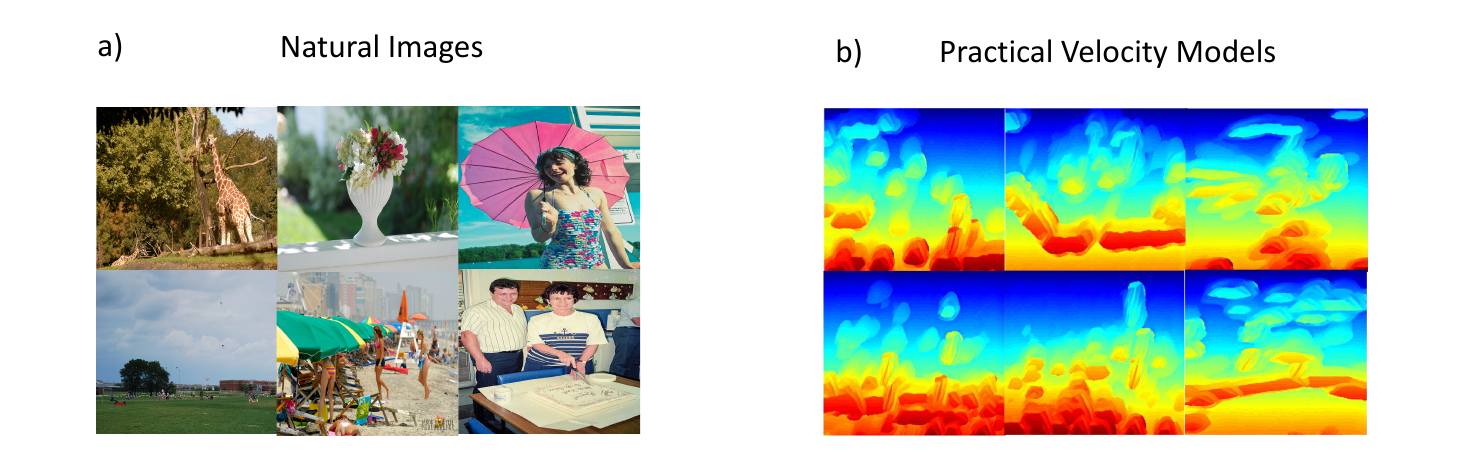}
\caption{a) Natural images from COCO dataset~\cite{lin2014microsoft} and b) the corresponding synthesized physically practical subsurface velocity models.}
\label{fig:COCO_velocity}
\end{figure*}
\subsection{Inversion Procedure}
To summarize, we provide the workflow for the implementation of our Multiscale InversionNet as the following 3 steps:
\begin{enumerate}[i.]
\item\textbf{Data Preparation}: Apply style transfer on the natural images to generated velocity perturbations. Combine the generated velocity perturbations with 1D velocity models to composite practical velocity models.
\item\textbf{Low-resolution Inversion}: Smooth the composite practical velocity models and generate their corresponding seismic data using forward modeling. Use the smoothed velocity models and their seismic data to train the Low-resolution InversionNet. Then apply the trained Low-resolution InversionNet on the test data.
\item\textbf{High-resolution Inversion}: Generate the seismic data with the high-resolution practical velocity models. Use the high-resolution velocity models and seismic data to train the High-resolution InversionNet. Then apply the trained High-resolution InversionNet on the test data.
\end{enumerate}

\section{Results}

\subsection{{Synthetic Tests}}
\subsubsection{Data Preparation}

We first conduct a synthetic test to demonstrate the performance of our methodology. 67,000 natural images from the COCO dataset~\cite{lin2014microsoft} are used as the content images and the Marmousi~\cite{brougois1990marmousi}
velocity model as the style image to construct the physically practical subsurface velocity models. The geometry of the Marmousi velocity model is based on a profile of the North Quenguela through the Cuanza basin~\cite{versteeg1993sensitivity}. The Marmousi model was built to resemble a continental drift geological setting. It contains many typical geological structures, such as interfaces, faults, and strong velocity variations in both the lateral and the vertical direction~\cite{brougois1990marmousi}.  Figure~\ref{fig:COCO_velocity} shows the natural images from the COCO data set and the generated practical velocity models using our approach~(as illustrated in Figure~\ref{fig:style_net}). In order to obtain velocity models with different resolutions, the practical velocity models are smoothed by a Gaussian filter with random standard deviation from 6 to 10 as low-resolution velocity models and smoothed with random deviation from 0 to 5 as high-resolution velocity models.

\begin{figure*}[ht]
\centering
\includegraphics[width=1.7\columnwidth]{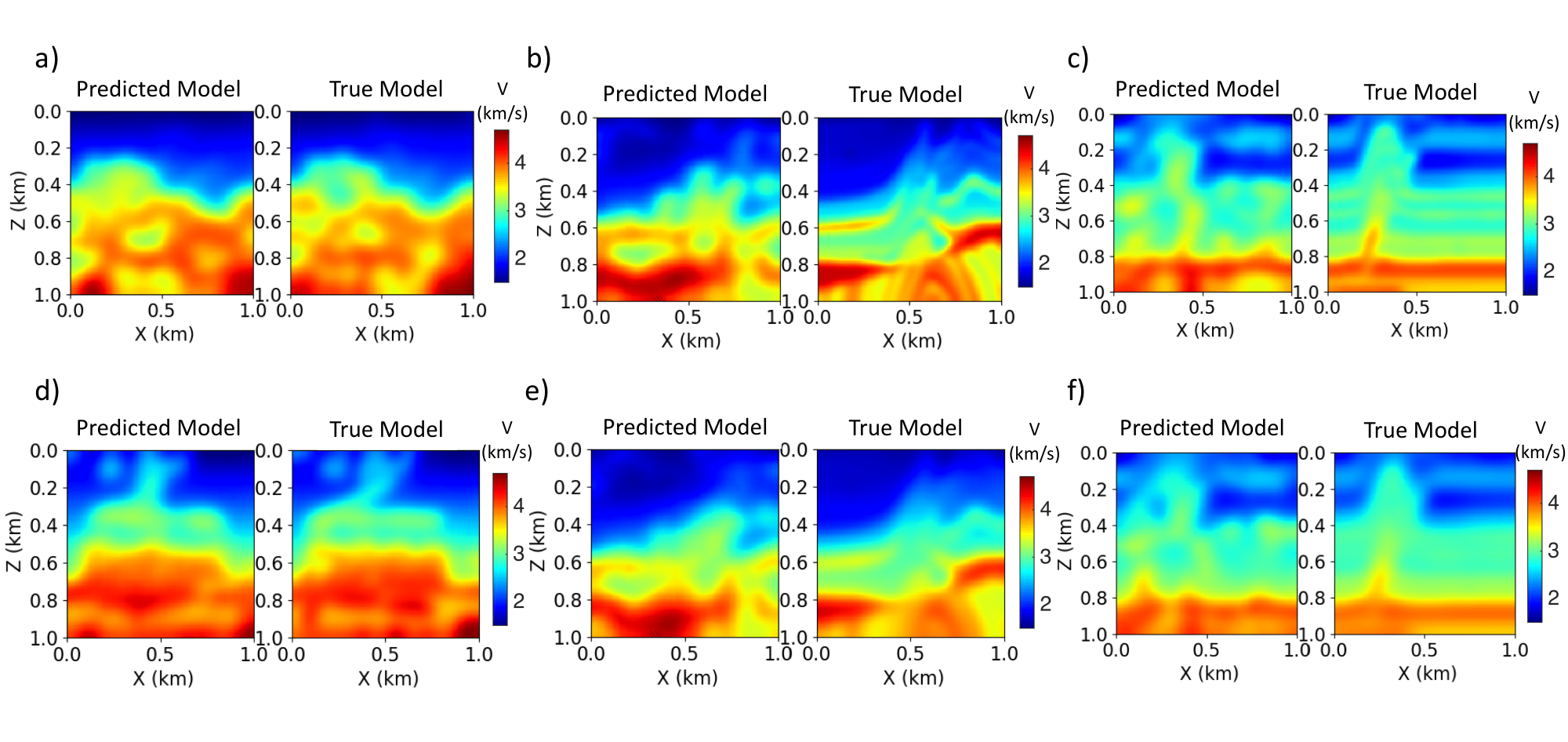}
\caption{Synthetic test results using Low-resolution InversionNet. a) and d) The predicted models and true models in the test set. b) and e) The predicted models and true Marmousi models with different resolutions. c) and f) The predicted models and true Overthrust models with different resolutions.}
\label{fig:mam_result}
\end{figure*}

\begin{table*}
\centering
\caption{The MSE and SSIM losses with the Low-resolution InversionNet results}
\begin{tabular}{c|cccccc} 
\hline
     & \begin{tabular}[c]{@{}c@{}}Test \\Model (a)\end{tabular} & \begin{tabular}[c]{@{}c@{}}~Marmousi\\~Model (b)\end{tabular} & \begin{tabular}[c]{@{}c@{}}~Overthrust \\Model (c)\end{tabular} & \begin{tabular}[c]{@{}c@{}}Test \\Model (d)\end{tabular} & \begin{tabular}[c]{@{}c@{}}Marmousi \\Model (e)\end{tabular} & \begin{tabular}[c]{@{}c@{}}Overthrust \\Model (f)\end{tabular}  \\ 
\hline
MSE  & 0.0212                                                   & 0.1065                                                        & 0.0343                                                          & 0.0138                                                   & 0.0720                                                       & 0.0200                                                          \\
SSIM & 0.7820                                                   & 0.5266                                                        & 0.6374                                                          & 0.7846                                                   & 0.6397                                                       & 0.7377                                                          \\
\hline
\end{tabular}
\label{table:low_inversion}
\end{table*}

These practical subsurface velocity models have been reshaped to a size of 2~km in both $x$ and $z$ directions with a grid spacing of 10~m. Ten sources are located on the surface with a spacing of 200 m, and the traces are recorded by 200 receivers spaced at an interval of 10 m. The source wavelet is a Ricker wavelet with a peak frequency of 15~Hz~\cite{wang2015frequencies}. We choose a Ricker wavelet as the source function to generate seismic waves due to its empirical success in processing seismic field data~\cite{fichtner2010full}. The seismic data are generated using the finite-difference method applied to an acoustic wave equation~\cite{alford1974accuracy}.

To validate the performance of our approach, we test our network on both in-distribution and out-of-distribution datasets. In particular, the in-distribution datasets are randomly selected from our test data~(as shown in Figures~\ref{fig:mam_result}(a) and (d)). Two out-of-distribution datasets, i.e., Marmousi and Overthrust~\cite{aminzadeh1996three} velocity models, are selected for testing due to their popularity~(as shown in Figures~\ref{fig:mam_result}(b) and (e), and Figures~\ref{fig:mam_result}(c) and (f), respectively). We smooth the Marmousi and Overthrust velocity models by a Gaussian filter with random standard derivation from 0 to 10 to validate if the low-resolution inversion can extract a low-resolution velocity model even if the data are generated by velocity models with different resolution.

\subsubsection{Low-resolution Inversion}
Similar to conventional multiscale full waveform inversion, the first step is to construct the low-frequency component of the velocity model from the data. To train the Low-resolution InversionNet, 65,000 pairs of low-resolution velocity models and their corresponding seismic measurements are used. An additional 2,000 pairs are used for testing. Every 10 examples is a training batch. Since the velocity models are smooth, the direct and transmission waves are dominated in the common-shot-gathers (CSGs)~\cite{reshef1986migration}.  To reduce the computation burden, the CSGs are down-sampled from $nz\times{nx}= 5000\times{200}$ to $200\times{200}$. 

We employ the Adam optimizer~\cite{kingma2014adam} to train the network with 50 epochs and the loss curve is given in the Supplemental Material. The results are compared in Figure~\ref{fig:mam_result} and their mean-square errors (MSE) and structural similarity indexes (SSIM)~\cite{hore2010image} are listed in Table~\ref{table:low_inversion}. We observe that the shallow parts of the velocity models are inverted well for all the tests, but there are some mismatches between the predicted models and true models in the deep areas. This is because transmission waves dominate the seismic data. 

\begin{figure*}[ht]
\centering
\includegraphics[width=1.7\columnwidth]{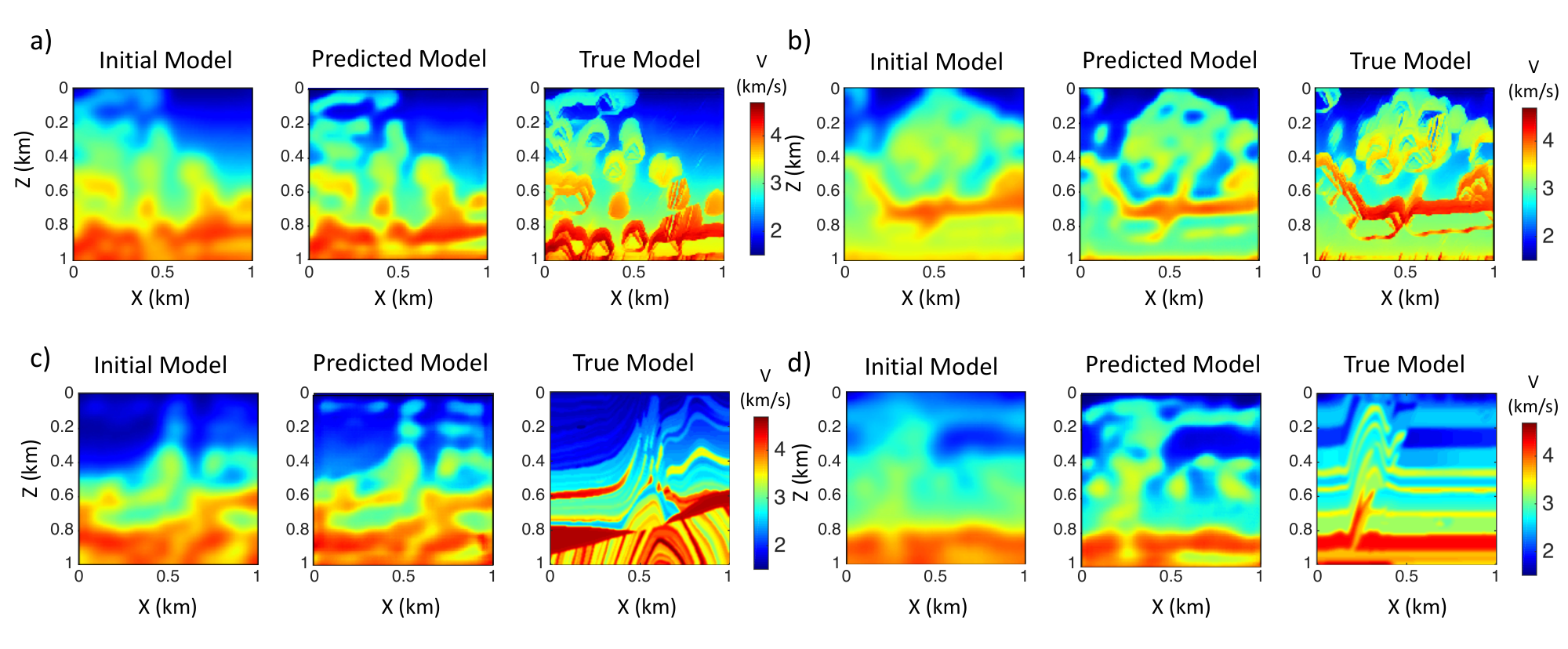}
\caption{Synthetic test results using High-resolution InversionNet. The initial models, predicted models and true models in a) test model 1 and b) test model 2. The initial models, predicted models and true models with c) Mamoursi model and d) Overthrust model.}
\label{fig:fcn_high}
\end{figure*}

\subsubsection{High-resolution Inversion}

Another neural network is built to construct the high-frequency components. Similar to conventional FWI approaches, the result from the low-resolution inversion is used as the initial guess. The data residual between the initial models and the true models is calculated. 65,000 groups of high-resolution velocity models, initial guesses and their data residual are used for training with High-resolution InversionNet and another 2,000 groups are used for testing, respectively. 

\begin{figure*}[ht]
\centering
\includegraphics[width=1.7\columnwidth]{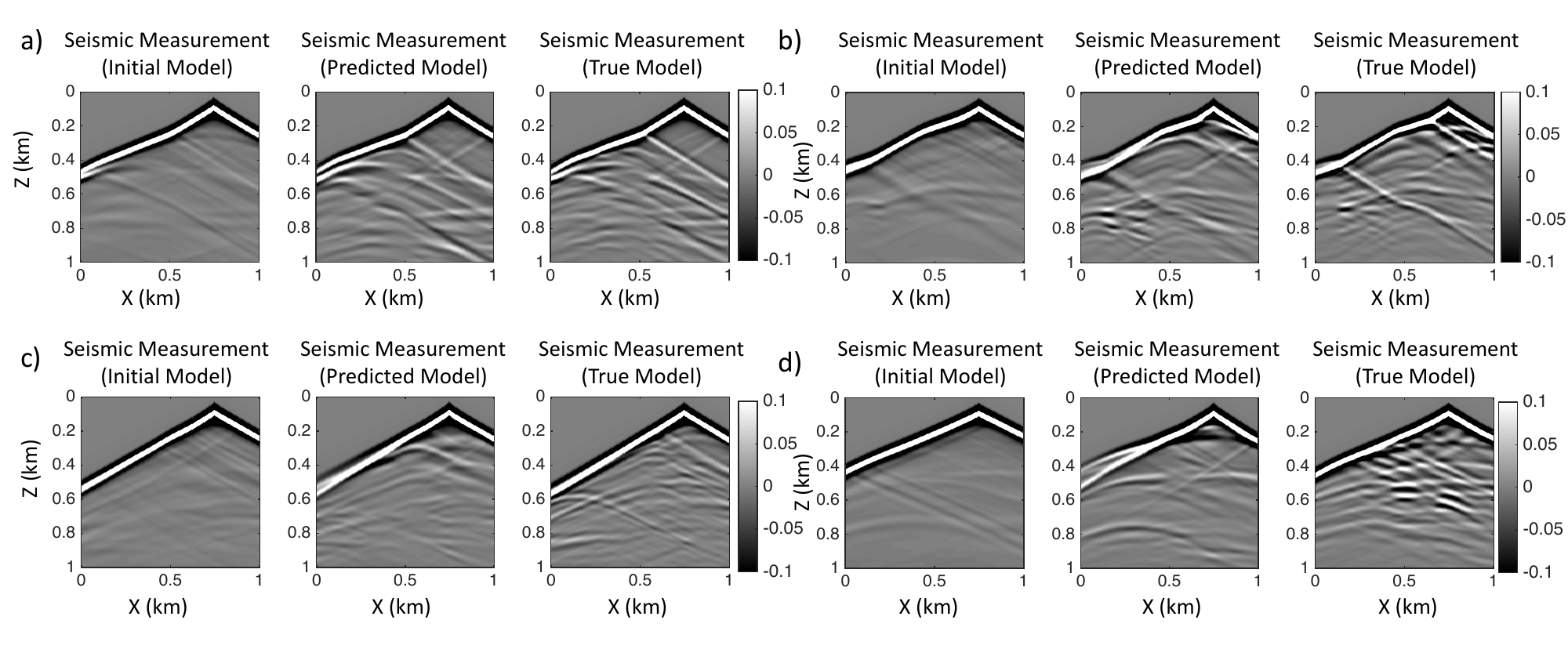}
\caption{The seismic measurements simulated with the initial models, predicted models and true models in a) test model 1 and b) test model 2. The initial models, predicted models and true models with c) Mamoursi model and d) Overthrust model given by High-resolution InversionNet.}
\label{fig:fcn_data2}
\end{figure*}

The High-resolution InversionNet is trained with 20 epochs and the loss curve is given in the Supplemental Material. Accordingly, we provide the reconstruction results in Figures~\ref{fig:fcn_high} and their MSE and SSIM losses in Table~\ref{table:high_inversion}, respectively.
We notice that the resolution of the velocity models has been significantly improved from the low-resolution results. However, the small reflection events in the models cannot affect the loss much since the loss function is based on the velocity models. As a consequence, only the big reflection events can be inverted clearly while the small events cannot be seen. Particularly, for the Marmousi test data, the shallow parts are inverted well. However, the folds are inverted as anomalies in the shallow parts since there are no bending structures in our training set, these anomalies leads to an increase in the $\ell_2$ loss of velocity models. For the Overthrust test data, the reflectors at $x= 0.2$~km and $0.8$~km are inverted correctly. However, the structures from $x= 0.2$~km to $0.5$~km are too complex so that the structure is strongly distorted. Moreover, the thin flat reflectors from $z = 0.4$~km to $0.8$~km are not inverted correctly because of the lack of such kind of structures in the training set. We have simulated the seismic measurements using the initial models and predicted models from the High-resolution InversionNet as shown in Figure~\ref{fig:fcn_data2}.


\begin{table*}
\centering
\caption{The MSE and SSIM losses with the High-resolution InversionNet result}
\begin{tabular}{c|cccc} 
\hline
     & \multicolumn{4}{c}{Initial Models}                                 \\
     & Test Model 1 & Test Model 2  & Marmousi Model & Overthrust Model  \\ 
\hline
MSE  & 0.0650        & 0.0741        & 0.1821           & 0.0709          \\
SSIM & 0.4867        & 0.4938        & 0.1693           & 0.3586          \\ 
\hline
     & \multicolumn{4}{c}{Predicted Models with L2 Loss}                   \\
     & Test Model 1  & Test Model 2  & Marmousi Model & Overthrust Model  \\ 
\hline
MSE  & 0.0361        & 0.0501        & 0.2281           & 0.1188          \\
SSIM & 0.5394        & 0.4950        & 0.1522           & 0.2895          \\ 
\hline
\end{tabular}
\label{table:high_inversion}
\end{table*}

\subsection{Field Data Test}
\subsubsection{Training Data Preparation and Field Data Description}

Employing inversion algorithms on test data is challenging for both physics-based and data-driven inversion methods. In this section, we test our method on a 2D Gulf of Mexico (GOM) data set~\cite{feng2017skeletonized,huang2018full} and compare it with two physics-based seismic inversion methods: wave equation traveltime tomography (WT)~\cite{luo1991wave} and multiscale FWI. The initial model for the multiscale FWI is obtained from traveltime tomography. 

We use the same 67,000 practical velocity models in the synthetic tests except the size of the velocity models are reshaped into 8.125~$km$ in the $x$ direction and 1.5~$km$ in the $z$ direction, with a grid spacing of 6.25~$m$. To make the survey configuration consistent with the field survey, 10 shot gathers are distributed on the surface of the model with a shot interval of 375 m, and each shot is recorded by a 6 km long cable with 480 receivers having a 12.5 m receiver interval. The shortest offset is 200 m. The source wavelet is extracted from the raw data by stacking the time-shifted reflection events together from 200 to 250 m offset in the shot gathers. 

\subsubsection{Low-resolution Inversion}

\begin{figure}[ht]
\centering
\includegraphics[width=0.9\columnwidth]{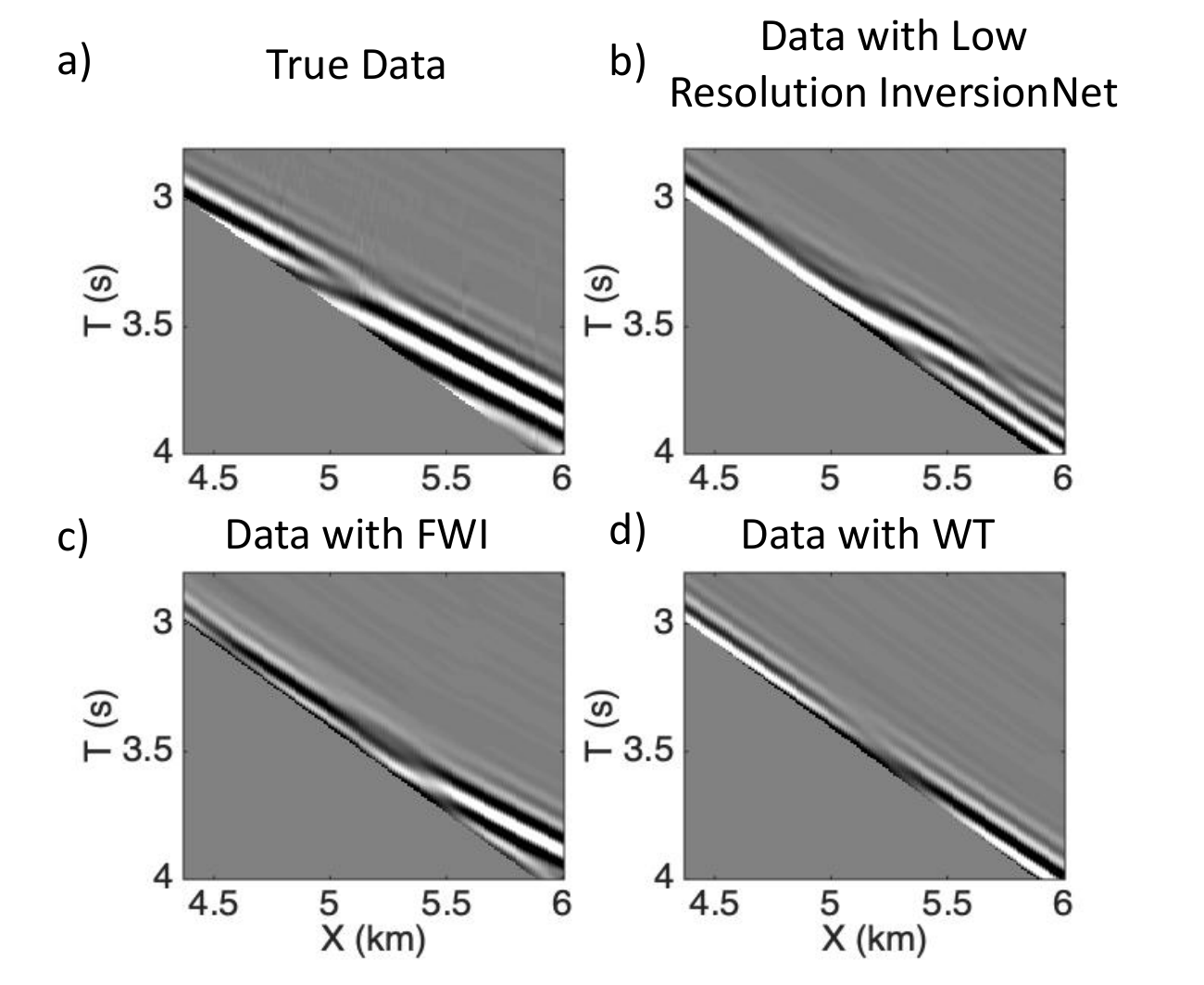}
\caption{a) True transmission data and data simulated with b) Low-resolution InversionNet result, c) FWI tomogram and d) WT tomogram.}
\label{fig:gom_data}
\end{figure}

To invert the low-frequency components of the velocity models, we generate seismic data using 65,000 low-resolution velocity structures. The reflection waves in the seismic measurements are muted, which means all the waves after the direct waves are removed~\cite{yilmaz2001seismic}. The CSGs are then downsampled to $200\times{200}$ for training. An example of the muted field data is shown in Figure~\ref{fig:gom_data}(a). We observe that the transmission waves are dominating and a small portion of reflection events remains in the data. The Low-resolution InversionNet has been trained for 10 epochs and then tested on the test data and GOM data. 

\begin{figure*}[ht]
\centering
\includegraphics[width=1.55\columnwidth]{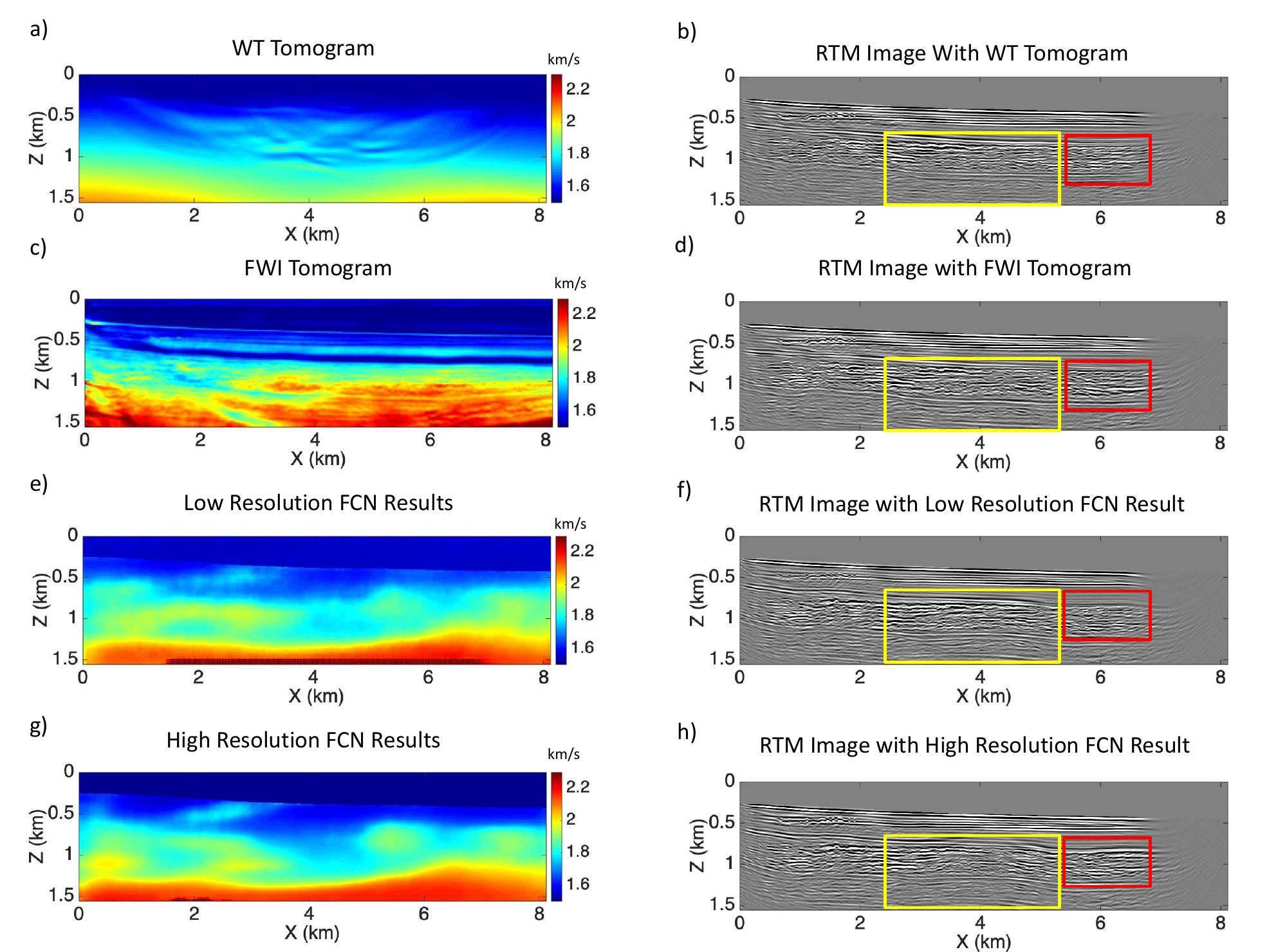}
\caption{a) and b) WT tomogram and its RTM image. c) and d) FWI tomogram and its RTM image. e) and f) Low-resolution InversionNet result and its RTM image. g) and h) High-resolution InversionNet result and its RTM image. }
\label{fig:gom_fcn}
\end{figure*}

\begin{figure*}[ht]
\centering
\includegraphics[width=1.6\columnwidth]{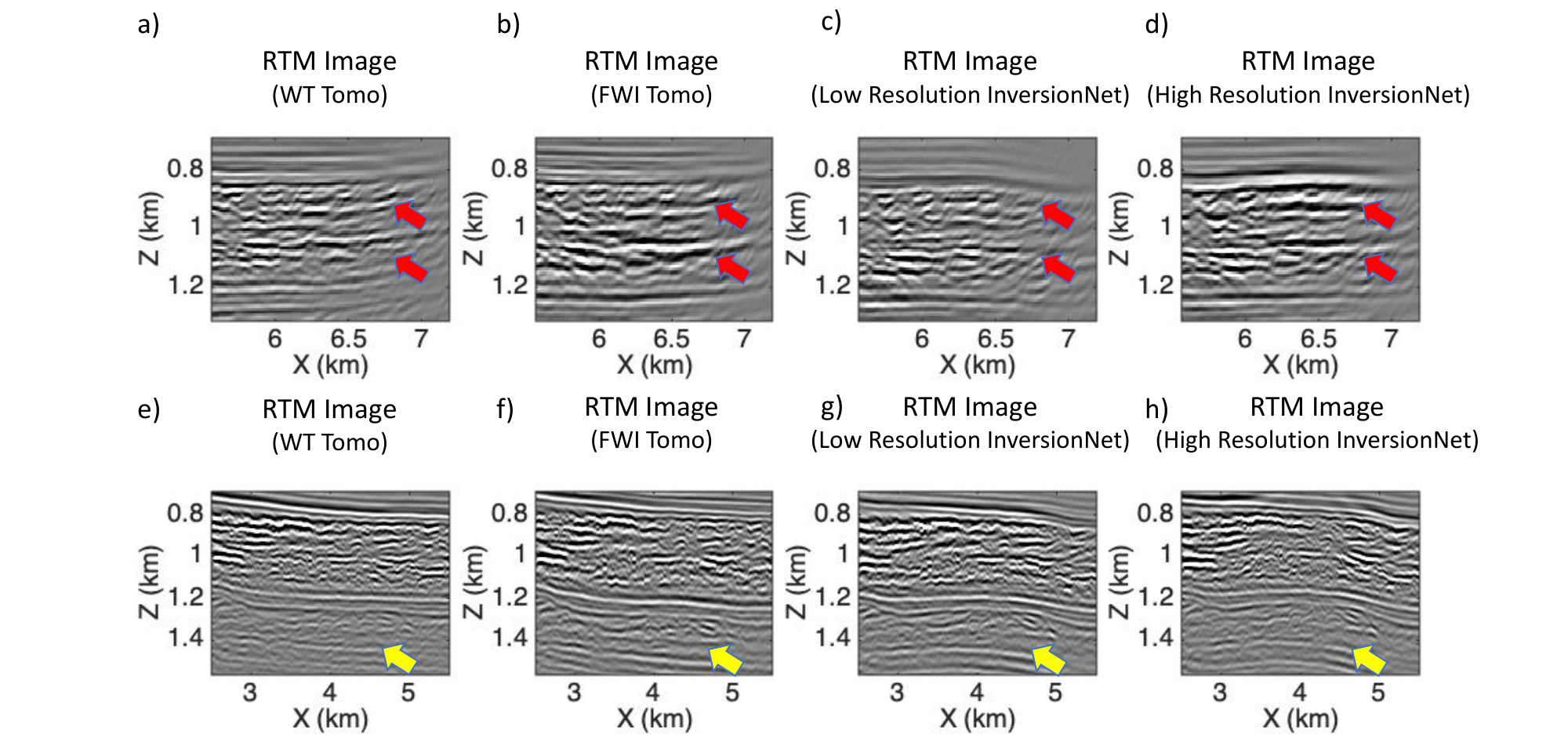}
\caption{The enlarged visualization of the RTM images in red boxes~(Row~1) and yellow boxes~(Row~2) in Figures~\ref{fig:gom_fcn}(b), \ref{fig:gom_fcn}(d), \ref{fig:gom_fcn}(f) and \ref{fig:gom_fcn}(h).}
\label{fig:gom_rtm_zoom}
\end{figure*}

Unlike tests using synthetic data, justifying the quality of inversion results from field data can be challenging in that it requires both qualitative tools and domain knowledge. In this work, we use reverse time migration~(RTM) as a quality control~(QC) tool to validate the inversion results from WT tomograms~\cite{feng2019transmission+} and full-waveform tomograms~\cite{huang2018full}. RTM is a seismic imaging technique, which provides the subsurface reflectivity using recorded seismic data. It is highly sensitivity to kinematic velocity errors, which result in a defocused and incoherent migration image~\cite{yang2019least}. Hence, RTM has been widely used as a QC tool for velocity model building~\cite{shan2008velocity,feng2019zero}. We provide the RTM images corresponding to different inversion results in Figure~\ref{fig:gom_fcn}. The enlarged visualizations of the yellow and red boxes in the RTM images are shown in Figure~\ref{fig:gom_rtm_zoom}. Since both WT inversion and Low-resolution InversionNet inversion mainly focus on the transmission waves, the RTM images in~\ref{fig:gom_fcn}(b) and \ref{fig:gom_fcn}(f) are comparable. However, there are a few reflection waves left in the muted data. Hence, the Low-resolution InversionNet can invert and obtain the velocity structure in the deep region. As a result, the reflection events in the deep regions, such as the reflection events with yellow arrows in Figure~\ref{fig:gom_rtm_zoom}(c) is more focused and continuous than that them in Figure~\ref{fig:gom_rtm_zoom}(a). 

\begin{figure}[ht]
\centering
\includegraphics[width=1\columnwidth]{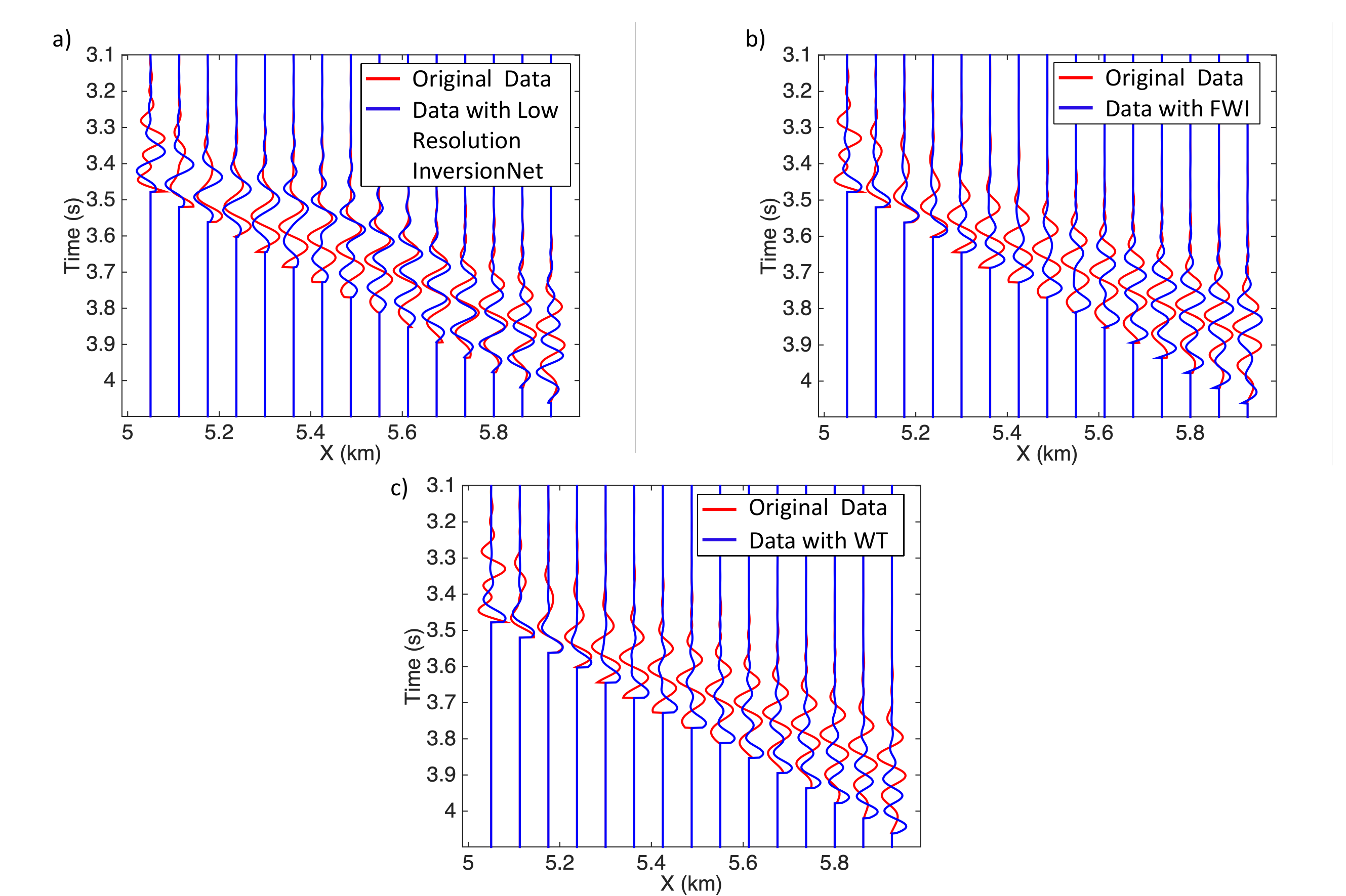}
\caption{ Wiggle trace comparison between original data and data simulated with a) Low-resolution InversionNet result, b) FWI tomogram and c) WT tomogram.}
\label{fig:gom_data_Wigg}
\end{figure}

In addition, we simulate transmission data with the predicted models using WT and FWI tomograms as shown in the Figure~\ref{fig:gom_data}. For comparison, these data are filtered by a 10~Hz low-pass filter. To better visualize the differences in data, we provide the traces in Figure~\ref{fig:gom_data_Wigg}. We observe that the data simulated from Low-resolution InversionNet~(Figure~\ref{fig:gom_data_Wigg}(a)) is reasonably consistent with the original data. The data from FWI~(Figure~\ref{fig:gom_data_Wigg}(b)) has a phase shift when compared to the original data, due probably to cycle-skipping. For the data with WT~(Figure~\ref{fig:gom_data_Wigg}(c)), the waveform does not match the original data since WT only considers the first arrival traveltime.

\subsubsection{High-resolution Inversion}
For inverting the high-frequency component of the velocity model, we make use of the reflection waves to update the velocity model from Low-resolution InversionNet. We mute all the waves before the direct waves and preserve reflection waves. After training for 5 epochs with the Low-resolution results as the initial models, we apply the High-resolution InversionNet on both the test data and GOM data.  RTM is used as the QC tool since the velocity errors leads to defocused and discontinued RTM images~\cite{shan2008velocity}. The enlarged visualizations of the RTM images in this area are shown in Figure~\ref{fig:gom_rtm_zoom}. Compared to the transmission waves, the illuminations of the reflection waves are deeper and wider. The boundary part of the velocity model from $X=5.5$ to $7.2$~km can be updated by the High-resolution InversionNet. Since both FWI and High-resolution InversionNet take advantages of the reflection wave, the RTM image as pointed by the red arrows in Figure~\ref{fig:gom_rtm_zoom}(b) and~\ref{fig:gom_rtm_zoom}(d) are more continuous and focused than those in Figure~\ref{fig:gom_rtm_zoom}(a) and~\ref{fig:gom_rtm_zoom}(c), which only utilize transmission waves. However, the reflection events, as pointed by the yellow arrows in Figure~\ref{fig:gom_rtm_zoom}(g), are more focused than those in Figure~\ref{fig:gom_rtm_zoom}(h). This is caused by the low-velocity zone at $X=3\sim5$ km and $Z=0.75\sim1.25$ km in Figure~15d. Since we use the low-resolution velocity model as the input and the high-resolution model as the output for the High-resolution InversionNet. The U-net boosts the high-resolution components regardless of the residual data. The low-velocity zone is reconstructed to a lower value than the true value.


Then we compare the FWI tomograms and High-resolution result in Figure~\ref{fig:gom_fcn}c and~\ref{fig:gom_fcn}g.
The velocity starts to increase around $Z=0.5$~km for both results, but the difference is the low-velocity zone from $X=3$ to $5$~km in the High-resolution InversionNet results. As a result, the enlarged visualizations of the RTM images in Figure~\ref{fig:gom_rtm_zoom}(f) and~\ref{fig:gom_rtm_zoom}(h) are different. But both of them are continuous, so it is hard to judge which one is better. It is hard to tell why there is such a big difference since the network is a black box.

\begin{figure*}[ht]
\centering
\includegraphics[width=1.2\columnwidth]{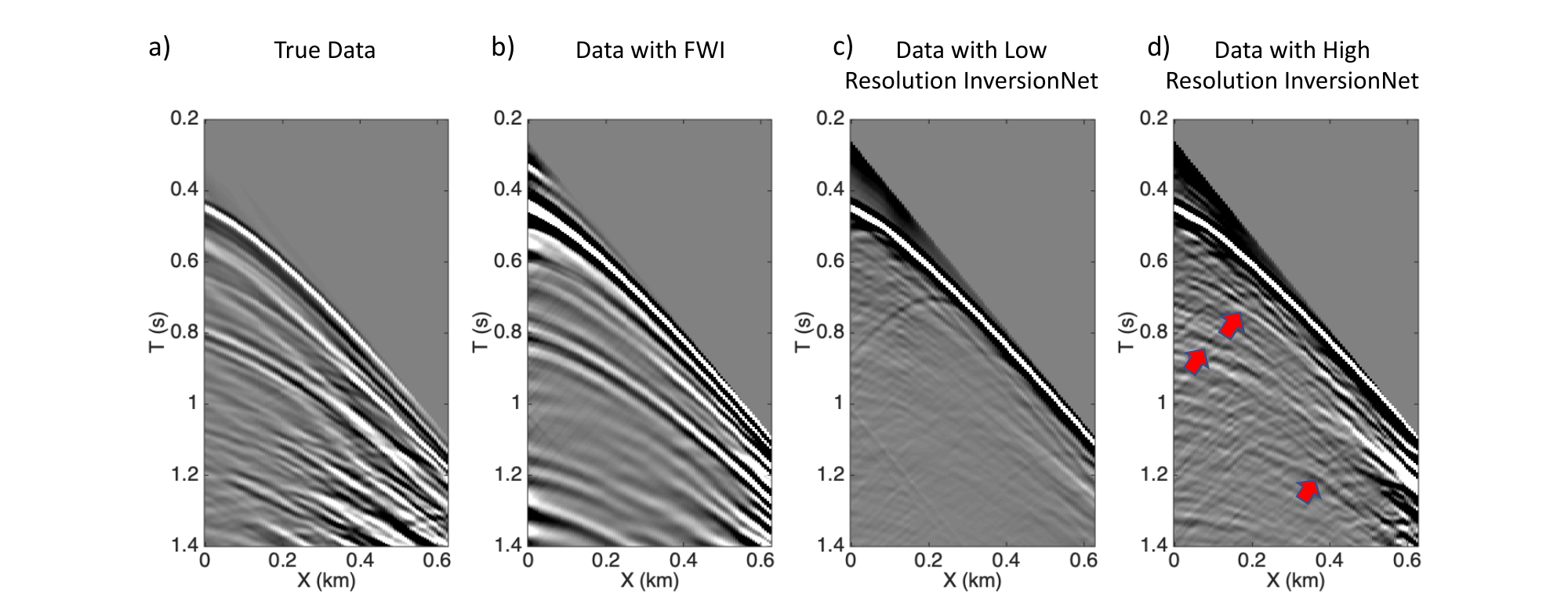}
\caption{a) True reflection data and data simulated with b) FWI tomogram, c) Low-resolution InversionNet result, and d) High-resolution InversionNet result. }
\label{fig:csg_refl_check}
\end{figure*}

Moreover, we compared the simulated reflection data of different inversion methods in Figure~\ref{fig:csg_refl_check}. Since the conventional FWI using data loss as the loss function, the reflection waves in Figure~\ref{fig:csg_refl_check}b are in general consistent with the true data in Figure~\ref{fig:csg_refl_check}a. For data with Low-resolution InversionNet, there are no reflection events due to the smooth velocity model. For the High-resolution InversionNet, only strong reflection events can be recovered. The reflection events that red arrows point in Figure~\ref{fig:csg_refl_check}d matches those in the true data.

\subsection{Additional Numerical Tests}

To help with understanding the performance of our proposed approaches, we carried out a few more numerical tests. Due to the page limit, we described those additional tests in the Supplemental Material to this paper. Those tests include: (1) a comparison of the performance using $\ell_1$ loss function versus $\ell_2$ loss function; (2) test of our strategy without using multiscale method; (3) a side-by-side comparison to physics-based FWI; (4) comparison to some of the existing velocity modeling methods; and (5) test on multiple input styles.

\section{Discussion And Future Work}

\subsection{``Doing More with Less''-- Incorporating Physics in Scientific Machine Learning}

Domain-aware learning is a unique and critical task to scientific machine learning~(SciML)~\cite{Basic-2019-DOE}. Our work explores particular avenues of incorporating critical physics knowledge in SciML through the problem of computational imaging. Labeled real data is extremely valuable but also costly to obtain in scientific applications. Our approach provides a feasible and cost-effective solution to address the dilemma of data scarcity for deep learning approaches. Built on style-transfer approaches, we develop a method to synthesize physically practical subsurface structure images that can be used to augment the training set and enrich the representativeness of the data. We have compared this method with several existing velocity model building methods in the Supplemental Material. Our approach has significant potential in that it not only leverages the existing large volume of natural images with  diversified representation, but also accounts for critical physics and domain knowledge. 

Many scientific problems involve systems that are governed by complex physical laws. It has recently been shown in the literature that there is a benefit to accounting for those physics in the design of the neural networks~\cite{Raissi-2019-Physics}. We propose and develop a particular strategy to decouple complex physics phenomena into simple ones, which can be separately incorporated into neural networks.  Comparing to those end-to-end strategy of incorporating domain knowledge, our approach of multiscale data-driven method better leverages the physics information, which results in  significantly improved imaging results with much higher resolution.

\subsection{Scientific ``Sim2Real''}

Physical simulation is an important tool for scientific problems. Originated in the robotics and vision community, the concept of ``Sim2Real'' refers to the ideas of transferring knowledge learned in simulation to the real data~\cite{Sim-2019-James}. Due to the lack of real labeled data in subsurface geophysics, model-based simulators have been widely used to synthesize simulations. However, pure model-based simulators usually simplifies the complex physics, which result in unavoidable reality gap between the simulation and real data. This gap degrades the predictivity and generalization ability of a predictive model. Our approach, on the other hand, is model-free and it learns the heuristic physics implicitly through the data without explicitly imposing physics rules. We demonstrate its capability in learning the physics to generate physically practical data for training predictive model. We further apply our predictive model to both out-of-distribution synthetic test data and real test data set. The results obtained are promising, which in turn proves the effectiveness of our approach in synthesizing simulation. However, there are still artifacts and missing information in the inversion results observable when applying our data-driven techniques to the Overthrust data set~(as shown in Figures~\ref{fig:fcn_high}. To further improve the inversion, a more diversified training set would be needed to capture various kinds of subsurface geology structures. One potential approach to increase the representativeness of our training set is to incorporate multiple subsurface style models with orthogonal features~(meaning styles represented by different geology images do not overlap). We have discussed the selection of style images and test InversionNet with multiple subsurface style models in the Supplemental Material. 

\subsection{Computational Benefits and Broader Applicability}

Our work is to address two important challenges in data-driven computational seismic imaging: accuracy and efficiency. As shown in our numerical tests, our data-driven inversion method outperforms the conventional FWI methods by alleviating several major computational issues such as local minima and need of good initial guess. We also show that once fully trained, our data-driven inversion model can be significantly more efficient in inferring subsurface structure than the conventional seismic imaging techniques. We demonstrate the efficacy of our model using both synthetic and field data.

Similar computational challenges exist among many computational imaging problems including seismic imaging. Although we demonstrate the performance of our techniques using computational seismic imaging problem, our methods are not restricted to this particular application. It can be potentially applicable to much broad computational imaging problems such as medical ultrasound tomography, radar imaging, microscope imaging, and many others. 

\subsection{Future Work}

Different approaches have been used in decoupling the complex wave phenomena. Our approach is one of the many. It would be worthwhile to explore the benefits of other means in decoupling complex waves. An example of this would the frequency decomposition, where a wave is decomposed into different frequency bands~\cite{Bunks-1995-Multiscale}. However, our network structure is designed to be technically flexible in incorporating various decompositions of wave physics.  

Loss function plays an important role in obtaining a predictive model. Throughout this work, we employ loss function on the subsurface velocity domain to justify the correctness of the inversion model. Our tests show that, that once converged, the training accuracy of our model can reach as high as $95\%$, which leads to a successful reconstruction of the major subsurface structures through training. However, it is the last $5\%$ of the training error that would contribute to further refining the subsurface structures with more details. This issue is essentially caused by the fact that our loss function lacks of data consistency. Similar problems have been also identified in other computational imaging applications~\cite{Otero-2020-Computed}. One potential approach to compensate for missing details would be a cycle-consistency loss~\cite{Unpaired-2017-Zhu}, which takes advantage of both the image loss as well as the data loss. This is one of our future direction. 

Physics-based regularization techniques have proved useful in improving resulting inversion for conventional FWI approaches. One example of those would be the illumination compensation, which have been usually used in conventional FWI methods to regularize the inversion and help with the deep regions~\cite{rickett2003illumination}. However, in our current model we have not yet employ any physics-based regularization to constrain our results. One of our future direction would be applying illumination regularization in the data domain along the $z$ direction for increasing the prediction accuracy.

Through numerical tests, we show that both the local features and global features play an important role in synthesizing practical velocity models. Our method has been demonstrated to be rather effective in learning the local features, however, it may not produce sufficient global features. Currently, our solution by using a simple 1D approximation on the background model can only provide limited global features. Extensive future work will be needed to address: (1) a more careful selection of the content natural images and (2) a different loss function in considering the global features.

\section{Conclusions}

We developed a multiscale data-driven seismic imaging technique. The inversion procedure has been separated into two steps that invert for low-frequency and high-frequency components of the velocity models, respectively. In particular, we design two different neural networks to account for different wave phenomena. To focus on the direct and transmission waves, we design the first network based  on InversioNet and train it with a smoothed velocity and their corresponding seismic measurements. To take the advantage of the reflection wave and refine the inversion results obtained from the first neural network, we further design and train the second neural network using high-resolution velocity models, the inversion results generated with the previous network and their data residuals. A high quality training set is the foundation for an effective data-driven inversion approach. We adopt a technique to generate physically meaningful subsurface velocity models with sufficient variability. Our technique is developed based on style transfer method that is capable of transferring a large amount of  natural images to practical subsurface velocity models. To validate the performance of our synthesized training set and the data-driven inversion techniques, we compare our approaches to conventional physics-based seismic imaging methods using both the synthetic and field data sets. Our results show that once fully trained using properly designed training set, our data-driven inversion model is much more efficient than those physics-based inversion methods and yields significantly improved imaging results.

\section*{Acknowledgements}

This work was co-funded by the U.S. DOE Office of Fossil Energy’s Carbon Storage program and by the Laboratory Directed Research and Development program of LANL under project numbers 20210542MFR and 20200061DR. We also thank Dr.~Tariq~Alkhalifah, an anonymous reviewer, and the Associate Editor for their constructive suggestions and comments that improved the quality of this work.

\bibliographystyle{IEEEtran}
\bibliography{main}

\end{document}